\documentclass[a4paper]{jpconf}

\usepackage{graphicx}
\usepackage[]{latexsym}
\usepackage{bm}

\newcommand{\be}{\begin{equation}}\newcommand{\ee}{\end{equation}}
\newcommand{\bea}{\begin{eqnarray}}\newcommand{\eea}{\end{eqnarray}}
\newcommand{\brr}{\begin{array}}\newcommand{\err}{\end{array}}
\newcommand{\bit}{\begin{itemize}}\newcommand{\eit}{\end{itemize}}
\newcommand{\ben}{\begin{enumerate}}\newcommand{\een}{\end{enumerate}}

\newcommand{\ba}{\begin{array}}
\newcommand{\ea}{\end{array}}

\def\lf{\left}

\def\ri{\right}

\def\1{{_{1}}}\def\2{{_{2}}}

\def\noHe0{:\;\!\!\;\!\!:H_e(0):\;\!\!\;\!\!:}
\def\noHm0{:\;\!\!\;\!\!:H_\mu(0):\;\!\!\;\!\!:}

\def\lf{\left}

\def\ri{\right}

\def\1{{_{1}}}\def\2{{_{2}}}

\begin{document}

\title{Geometric invariants as detector of Hawking and Unruh effects and quantum field theory in curved space}

\author{ Antonio Capolupo$^1$ and Giuseppe  Vitiello$^2$}
\address{$^1$Dipartimento di Ingegneria Industriale,
  Universit\'a di Salerno, Fisciano (SA) - 84084, Italy}
\address{$^2$Dipartimento di Fisica E.R.Caianiello
  Universit\'a di Salerno, and INFN Gruppo collegato di Salerno, Fisciano (SA) - 84084, Italy}

\begin{abstract}
We report on the recent results revealing the presence of geometric invariants in all the phenomena in which vacuum condensates appear and we show that Aharonov--Anandan phase can be used to provide the evidence of phenomena like Hawking and Unruh effects and to test some behavior of quantum field theory in curved space. A very precise quantum thermometer can be also built by using geometric invariants.
\end{abstract}

\section{Introduction}

Disparate physical systems show the presence of vacuum condensate and
are   effectively described in a similar way by Bogoliubov transformations.
Examples are represented by gravitational  effects like Unruh \cite{Unruh:1976db}, Hawking \cite{Hawking:1974sw} and Parker \cite{Parker:1968mv} ones, phenomena like Schwinger effect \cite{Schwinger:1951nm}, superconductivity \cite{Bardeen:1957mv}, Thermo Field Dynamics \cite{Takahasi:1974zn}, particle mixing \cite{Blasone:1995zc,Capolupo:2006et,Capolupo:2010ek}, quantum dissipative systems \cite{Celeghini:1991yv} and graphene physics \cite{Iorio:2010pv}.
The formal differences among these systems are contained in the Bogoliubov coefficients.

Some of these phenomena, like Hawking, Unruh and Parker effects, which consist in
 the nonperturbetive production of particle-antiparticle pairs from vacuum, are very hard to be detected
 and intensive study has been devoted to the tentative to observe such phenomena in analogous atomic systems and  to test quantum field theory in curved space-times in laboratory.

An apparently separated research line is represented by the study of geometric phases \cite{Berry:1984jv}--\cite{Capolupo:2011rd}
 appearing in the evolution of many physical systems.
 Such phases have been observed in a variety of systems \cite{Tomita,Jones,Leek,Pechal} and their potential applications in particle physics has been also studied \cite{Capolupo:2011rd}.

In the present paper we report the results of Ref.\cite{Capolupo2013} where the presence of geometric phase in all the systems effectively described by a Bogoliubov transformation has been unveiled. We show that Aharonov--Anandan (A-A) invariant can be used to prove experimentally, in atomic systems, the existence of Hawking radiation and Unruh effect, and to reveal some aspects of quantum field theory in curved background in particular graphene morphologies. A-A invariant can be also used to build a very precise thermometer.

The paper is structured as follows.
In Sec.II we recall some basic facts about Bogoliubov transformations and A-A invariant and we derive the expression for A-A phase for systems underlying to a Bogoliubov transformation.
In Sec.III we present the geometric invariant arising in thermal states and study its application to the Hawking effect in acoustic black hole. The application of A-A invariant to graphene physics and to the study of Unruh effect are presented in Secs.IV and V, respectively. The possibility to build a quantum thermometer by using A-A phase is shown in Sec.VI and Sec.VII is devoted to conclusions.

\section{Aharonov--Anandan invariant and vacuum condensate}

The Bogoliubov transformation \cite{Takahasi:1974zn} has the form:
$
\tilde{\alpha}^r_{\mathbf{k}}(\xi, t) = U^{\psi}_{\mathbf{k}} \, \alpha^r_{\mathbf{k}}(t) + V^{\psi}_{-\mathbf{k}} \, \alpha^{r\dagger}_{-\mathbf{k}}( t)\,.
$
It is required to be a canonical transformation, i.e. it must leave the canonical commutation relations (CCRs) invariant. This request translates in the condition $|U_{\mathbf{k}} |^2 \pm |V_{\mathbf{k}} |^2=1$, with $+$ for fermions and $-$ for bosons.
The Bogoliubov transformation  can be rewritten in terms of the generator $J(\xi, t)$ as:
$
\tilde{\alpha}^r_{\mathbf{k}}( \xi, t) = J^{-1} (\xi,  t)\,\alpha^r_{\mathbf{k}}(t) J(\xi,  t)\,.
$
In the case of the phenomena described by Bogoliubv transformations, the states physically relevant are the states  $|\widetilde{\psi}(\xi,  t)\rangle$ related to the original states $|\psi(t)\rangle$ by the transformation $|\widetilde{\psi}(\xi,  t)\rangle = J^{-1} (\xi,  t)|\psi(t)\rangle $.
For these states the variance of the energy is always different from zero. This fact implies that  A-A invariant appears in a similar way in all the phenomena representable by a Bogoliubov transformation.
Indeed, in order to generate the A-A phase \cite{Anandan:1990fq} in the time evolution of an isolated system  it is necessary and sufficient that the state of the system $|\phi(t)\rangle$  is not a stationary state, i.e. it has a nonzero value of the uncertainty $\Delta E(t) $  in energy,
$
\Delta E ^{2}(t) = \langle \phi(t)|H^{2}|\phi(t)\rangle -  (\langle \phi(t)|H|\phi(t)\rangle)^{2}.
$
 The A-A invariant is then defined as
$
s= \frac{2}{\hbar} \int_{0}^{  t}  \Delta E (t^{\prime}) \, dt^{\prime}\,.
$

In the case of  phenomena in Refs.\cite{Unruh:1976db}--\cite{Celeghini:1991yv}, the geometric phase, expressed without making explicit the expressions of the Bogoliubov coefficients, is formally the same for all the systems. In particular, for neutral scalar field, it is given by
$
S_{B}(t)=2\sqrt{2}\int_{0}^{  t} \omega_{\bf k} U^B_{\bf k}(t^{\prime}) V^B_{\bf k} (t^{\prime})dt^{\prime}.
$
Similar relation holds for fermions.
 We now study different specific cases.

\section{Aharonov--Anandan invariant and Hawking effect}

{\it Thermal state} -- Let us start by considering the thermal state defined in Thermo Field Dynamics \cite{Takahasi:1974zn}. The  parameter  $\xi$ characteristic of the Bogoliubov transformation describing the relation between the thermal and the non-thermal state depends on temperature. The relevant vacuum is the finite temperature one, and  the coefficients $U$ and $V$ for bosons are
$U^{T}_{\bf k } = \sqrt{e^{\beta\hbar \omega_{\bf k} }/(e^{\beta\hbar \omega_{\bf k } }-1)}$
and $V^{T}_{\bf{k}} = \sqrt{1/(e^{\beta\hbar \omega_{\bf k} }-1)}$, respectively, with $\beta = 1/ k_{B}T$.
Similar coefficients are found for fermions.
The Aharonov--Anandan invariant is
$
S_{T}(t) = 2\sqrt{2} \omega_{\bf k} t \,e^{\beta \hbar \omega_{\bf k}/2 }/(e^{\beta \hbar \omega_{\bf k } }-1)\,,
$
then, the phase difference
existing between two thermal states at different temperature  is
 \bea\label{DeltaphaseTermal}
\Delta S_{T}(t) = 2\sqrt{2} \omega_{\bf k} t \lf[\frac{e^{ \hbar \omega_{\bf k}/2 k_{B} T_{1} }}{e^{ \hbar \omega_{\bf k }/  k_{B}T_{1} }-1}-
\frac{e^{ \hbar \omega_{\bf k}/2 k_{B}T_{2} }}{e^{ \hbar \omega_{\bf k }/  k_{B}T_{2} }-1}\ri]\,.
\eea

{\it Hawking effect} -- An example in which the state  $|\widetilde{0}(\xi,  t)\rangle$ represents a thermal vacuum
is given by the Hawking effect. Then the invariant (\ref{DeltaphaseTermal}) can help to detect the Hawking radiation.

Such an effect consists in black body radiation that is predicted to be emitted by black holes, due to quantum effects near the event horizon. The thermal bath observed outside the event horizon of a black hole has a temperature depending on the black hole mass $M$, $T_{H}=\hbar c^{3}/8 \pi G M k_{B}$, where $G$ is the gravitational constant.
This effect can be described by a thermal state with temperature  $T_{H}$.

Recently,  an acoustic black hole  has been generated by accelerating  a Bose-Einstein condensate of $10^{5}$ atoms of $\;^{87}$Rb   to velocities which  exceed the sound velocity \cite{Lahav}. The sonic event horizon is represented by the point where the flow velocity equals the speed of sound. Thus sound waves, rather than light waves, cannot escape the event horizon.
 In this experiment the flow velocity is in only one direction.
 The effective temperature of Hawking radiation for such a system is given by $T_{H} = \hbar g_{H} / 2 \pi k_{B} c_{H} $, where $c_{H}$ is the sound speed at the black hole horizon, and  $g_{H}$ represents the effective surface gravity \cite{Lahav}.
 Such a temperature
  can be also written as $T_{H}=\mu / k_{B} \pi \lambda$,
 where $\mu = m c^{2}$ is the chemical potential of the condensate with $m$ atomic mass, $c$ speed of sound in the condensate, and  $\lambda$ represents a parameter corresponding to the number of correlation lengths need to have a thermal spectrum for the Hawking radiation. $\lambda = 7$ in \cite{Lahav}, moreover, the temperature of the acoustic black hole is $T_{H}\in (2-10) n K$. Since the temperature of the condensate $T_{cond}$ is $(20-170)nK$, the Hawking radiation is indistinguishable from the thermal noise and very difficult to identify also in such a system. An interferometric study could then be crucial in the detection of Hawking radiation in acoustic black holes. Indeed, a geometric phase similar to the thermal one is associated to Hawking radiation  with $T\equiv T_{H}$.

Such phase could be detected  via interferometry, measuring the difference of geometric phase
 between two flows of one-dimensional Bose-Einstein condensate, one in which is realized an acoustic horizon and the other in which the stream is subsonic. Such a configuration can be obtained by using two devices like the one  presented in \cite{Lahav}. In this way the Hawking radiation can be distinguished  from thermal noise,  by means of the presence of
  a difference of geometric phases $\Delta S_{H}$  between the supersonic and the subsonic condensate. $\Delta S_{H}$  is given by Eq.(\ref{DeltaphaseTermal}) where the relevant temperatures are  $T_{1}\equiv T_{H}+T_{cond}$ and $T_{2}\equiv T_{cond} $.

 By considering the limit imposed by the dimension of the acoustic black hole \cite{Lahav,cirac}  and the fact that the black hole horizon is maintained for about $20 ms$ in the experiments \cite{Lahav}, we plot  in Fig.1  $\Delta S_{H}$ as function of excitation energy for different black hole temperatures. Such pictures show that A-A phases are in principle detectable, thus they can be used to reveal Hawking effect in Bose-Einstein condensate.
\begin{figure}
\centering
\begin{picture}(300,180)(0,0)
\put(10,20){\resizebox{9 cm}{!}{\includegraphics{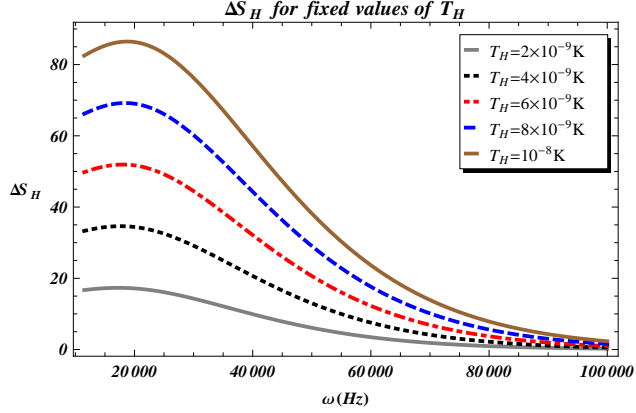}}}
\end{picture}\vspace{-1cm}
\caption{\em Plots of $\Delta S_{H}$
as a function of excitation energy $\omega$, for a time interval $t = 20 ms $ and for sample values of  $T_{H} \in  [2, 10] nK$, as indicated in the inset and $T_{cond}=40nK$.}
\label{pdf}
\end{figure}

\section{Aharonov--Anandan invariant and graphene}

 A system in which the presence of  A-A invariant can help to verify quantum field theory in curved space in table top experiment is represented by graphene. Recent studies on graphene have  shown that morphology of the sample
modifies its electronic properties. In  Ref.\cite{iorio2012} it has be shown that a sheet of graphene shaped as Beltrami pseudosphere, $dl^{2}=du^{2} + r^{2} e^{2 u / r}dv^{2}$, with $v\in [0,2\pi] $, $u\in [-\infty,0]$ and $r$ radius of curvature, displays a finite temperature electronic local density of states,
 thus revealing characteristic properties of quantum field theory in curved spaces. The temperature found in \cite{iorio2012} is $T = T_{0}e^{ u / r}$ with $T_{0} = \hbar v_{F}/(2 \pi k_{B}r)$ where $v_{F}$ is the Fermi velocity.
In \cite{iorio2012} an experiment with a Scanning Tunneling Microscope (STM) has been also proposed to detect the effect there  provided.
Here we show that such effect could be revealed easily by using the A-A invariant.

By using the local Weyl symmetry and the formalism presented in  \cite{iorio2012}, we find that the geometric phase
in graphene shaped as Beltrami pseudosphere is given by
\bea\label{graphene}
S_{g}(t) = 2\sqrt{2} \omega_{\bf k} t \frac{e^{ \hbar \omega_{\bf k}/2 k_{B}T }e^{- 2 u/r }}{e^{ \hbar \omega_{\bf k }/  k_{B}T }-1}\,,
\eea
Since the zero-curvature limit of finite temperature electronic local density of states does not match the flat electronic local density of states \cite{iorio2012}, for $r\rightarrow \infty$, Eq.~(\ref{graphene}) does not match the geometric phase in the case of flat space.

A phase difference of the tunneling current along two different meridians $u$ and $u^{\prime}$ can be measured
by fixing the energy $E=\hbar \omega$ and the curvature $r$ of the surface. Then, from the value of $\Delta S_{g}$ one derives the value of the temperature  $T_{0}$.
In Fig. 2 we plot $\Delta S_{g}$ as function of $T_{0}$ for values of the electron energy shown in the inset and for $u=-r/2$ and $u^{\prime}=-r/3$, with variable $r$.
\begin{figure}
\centering
\begin{picture}(300,180)(0,0)
\put(10,20){\resizebox{9 cm}{!}{\includegraphics{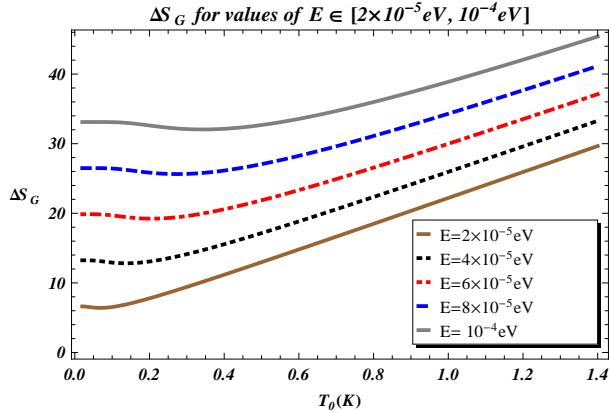}}}
\end{picture}\vspace{-1cm}
\caption{\em Plots of $\Delta S_{G}$ {\it vs} $T_{0}$, for a time interval $t = 10^{-10}s $ and for sample values of  $E \in  [2 \times 10^{-5} eV,    10^{-4} eV]$, as indicated in the inset.}
\label{pdf}
\end{figure}
We see that $\Delta S_{g}\neq 0$ in graphene for a wide range of values of $T_0$.

\section{Aharonov--Anandan and Unruh effect}

Another example in which the state  $|\widetilde{0}(\xi,  t)\rangle$ represents a thermal vacuum
is given by the Unruh effect.
In this phenomenon, the ground state for an inertial observer is seen as in thermodynamic equilibrium with a temperature different from zero  by the uniformly accelerated observer.
The Bogoliubov coefficients allow to express the Minkowski vacuum in terms of Rindler states. The temperature of thermal bath depends on the acceleration $a$ of the observer, $T_{U}=\hbar a/2 \pi c k_{B}  $.
The detection of such phenomenon is very hard, thus, also in this case, the geometric phase (\ref{DeltaphaseTermal}) could represent an useful instrument.

 Since, in free field case, the inertial systems has no A-A phase,
the geometric phase difference between the accelerated and inertial systems coincides with the phase of the accelerated system and Eq.(\ref{DeltaphaseTermal}) reduces to
$
\Delta S_{U}(t) = 2\sqrt{2} \omega_{\bf k} t \lf[e^{ \pi \omega_{\bf k} c / a }/(e^{ 2\pi \omega_{\bf k} c / a  }-1)\ri].
$
Such a equation describes completely the geometric phase produced in the gravitational Unruh phenomenon.
In this regard, let us note that many proposal have been suggested to  detect Unruh effect in atomic device  by means of Berry phase \cite{Hu-Yu,Ivette1}. It has be shown that such a phenomenon could be demonstrate by analyzing the phase variation due to the acceleration of a two level atom, which can be observed through interference with an inertial atom.

We note that the Berry phase  is defined only for systems which have an adiabatic and cyclic evolution. Here, we compute the A-A phase that is independent by the particular time evolution of the system and generalize the Berry phase to the noncyclic case. Thus   A-A  invariant can be more useful than Berry phase in the study of Unruh radiation. We consider the interacting Hamiltonian describing the interaction between the atom and the electromagnetic field in the multipolar scheme, $H = (1/2)\,\hbar\,\omega_{0}\,\sigma_{3}\,+\,H_{\phi}\,-\,e \sum_{mn}\mathbf{r}_{mn}\cdot \mathbf{E}(x(t))\sigma_{mn}$,   where $\omega_{0}$ is the energy level spacing of the atom, $H_{\phi}$ is the Hamiltonian of the electromagnetic field, and  $\mathbf{E}$ is the electric field strength. Assuming a weak interaction between atom and field, one has  the effective hamiltonian $H_{eff}=(1/2)\,\hbar\, \Omega\, \sigma_{3}$, with $\Omega$ renormalized energy level spacing which contains the Lamb shift term. Considering a  two level atom state, $|\phi(t)\rangle = \sin (\theta(t)/2)|+\rangle + \cos (\theta(t)/2)|-\rangle $ (with $\theta(t)$ defined in \cite{Capolupo2013,Hu-Yu}),
 the A-A invariant generated in the evolution of this state is $S = \, \Omega\,\int_{0}^{\tau} \sin  \theta(t) d \tau$.
For a two-level atom uniformly accelerated in the $x$ direction with acceleration $a$, the phase $S_{a}$ is
\bea\label{faseAA}
S_{a}\,=\pm\,\int_{0}^{\tau}\frac{e^{2 A_{a}t}\sin \theta}{\sqrt{e^{4 A_{a}t}\sin^{2} \theta\,+\,\lf(R\,-\,R\,e^{4 A_{a}t}\,+\,\cos^{2} \theta \ri)^{2}}} \,\Omega_{a}\, dt\,,
\eea
where $R_{a} = B_{a}/A_{a}$, with
$
A_{a}\,=\,(1/4)\,\gamma_{0}\lf(1+a^{2}/c^{2} \omega_{0}^{2} \ri)(e^{2 \pi c \omega_{0}/a}+1)/(e^{2 \pi c \omega_{0}/a}-1)  \,,$
 and
$ B_{a}\,=\,(1/4)\,\gamma_{0}\lf(1+a^{2}/c^{2} \omega_{0}^{2} \ri)\,,
$
with $\gamma_{0}\,=\,e^{2}\,|\langle -|\mathbf{r}|+\rangle|^{2} \omega^{3}_{0}/3\pi\varepsilon_{0}\hbar c^{3}$ spontaneous emission rate
and $\Omega_{a}$ effective level spacing of the atoms \cite{Capolupo2013,Hu-Yu}.

In the case of an inertial atom, $a=0$, the phase $S_{a=0}$ assume the same form of Eq.(\ref{faseAA}), with $A_{a}$, $B_{a}$, $R_{a}$ replaced by $A_{0}=B_{0}=(1/4)\,\gamma_{0}$ and  $R_{0}=1$, respectively.
The phase difference between the phases of accelerated and inertial atoms, $\Delta S_{U} = S_{a }-S_{a=0}$, gives the geometric phase due purely to the atom acceleration.  We neglect the very small contribution given by the Lamb shift term, $\Omega_{a} \sim \omega_{0}$,  and assume  $|\langle -|\mathbf{r}|+\rangle|\sim a_{0}$, where $a_{0}$ is the Bohr radius,  and $\omega_{0} \approx -e^{2}/8 \pi \hbar \varepsilon _{0} a_{0}$, so that $\gamma_{0}/\omega_{0} \approx 10^{-6}$ \cite{Hu-Yu}. By considering an initial state with angle $\theta = \pi /5$ and    $\omega_{0} \sim 10^{9} s^{-1}$, we obtain detectable phase differences  for    $a \in (10^{17}-10^{18}) m/s^{2}$, as shown in Fig.3.
These values of the accelerations are much less than the one required to detect Unruh radiation $(\sim 10^{26}m/s^{2})$ and are physically accessible with current technology.
\begin{figure}
\centering
\begin{picture}(300,180)(0,0)
\put(10,20){\resizebox{9 cm}{!}{\includegraphics{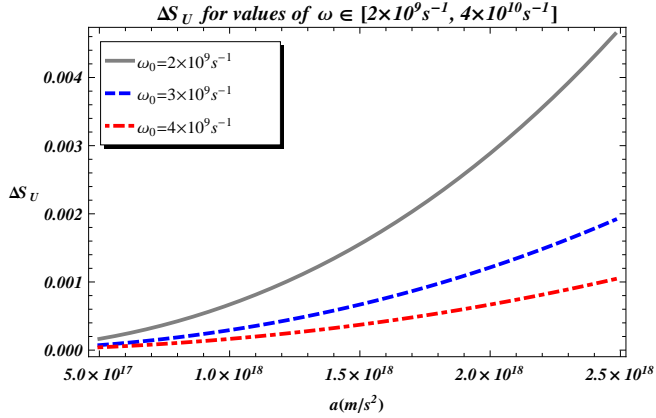}}}
\end{picture}\vspace{-1cm}
\caption{\em Plots of $\Delta S_{U}$
as a function of acceleration $a$, for a time interval $t = 4 \times 2\pi /\omega_{0} $ and for sample values of  $\omega_{0} \in  [2 \times 10^{9} s^{-1},  4 \times  10^{9} s^{-1}]$, as indicated in the inset.}
\label{pdf}
\end{figure}

\section{Quantum thermometer}

A further application of Aharonov--Anandan phase is represented by the possibility
to obtain very precise measurement of temperature by means of phase difference of two atomic systems at different temperatures.
Indeed, a  phase similar to the one in Eq.(\ref{faseAA}) is acquired by the atom  also if it interacts with a thermal state.
In this case the coefficients $A_{a}$ and $B_{a}$ are replaced by $A_{T}\,=\,(1/4)\,\gamma_{0}(1+4\pi^{2} k_{B}^{2}T^{2}/\hbar^{2} \omega_{0}^{2} )\,(e^{E_{0}/k_{B} T}+1)/(e^{E_{0}/k_{B} T}-1) \,,$
 and
$ B_{T}\,=\,(1/4)\,\gamma_{0}(1+ 4\pi^{2} k_{B}^{2}T^{2}/\hbar^{2} \omega_{0}^{2} )\,,
$ respectively, with $E_{0} = \hbar \omega_{0}$.

Thus, an atomic interferometer in which a single atom follows two different paths and interacts with two thermal states at different temperatures can represent a new type of very precise thermometer. The difference of geometric invariants between the two paths permits to determine the temperature of one sample once known the temperature of the other sample.
If it is known the temperature $T_h$ of the hotter source, for fixed values of $\omega_{0} $ and of time, one can  measure  the temperature $T_c$ of the colder cavity.
Considering the atomic transition frequencies  $\omega_{0}$, values of $T_h$  reported in Fig.4, and time intervals of order of
$t = 4 \times 2 \pi/\omega \,\, s$, we obtain measurement of temperatures of cold sources of about $2$ orders of magnitude below the reference temperature of the hot source, see Fig.4.
\begin{figure}
\centering
\begin{picture}(300,180)(0,0)
\put(10,20){\resizebox{9 cm}{!}{\includegraphics{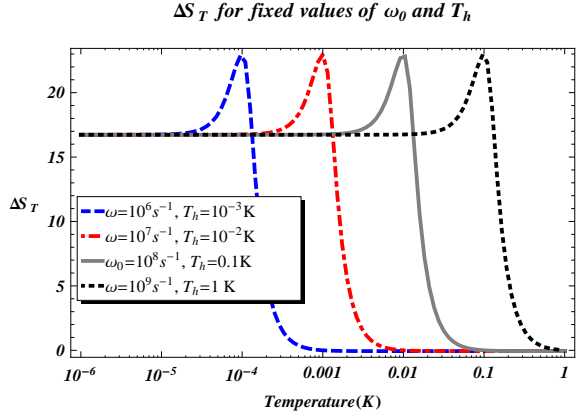}}}
\end{picture}\vspace{-1cm}
\caption{\em Plots of $\Delta S_{T}$
as a function of temperatures of cold sources  $T_{c}$, for sample values of  $\omega \in  [  10^{6} s^{-1}- 10^{9}  s^{-1}]$, time intervals $t = 4 \times 2 \pi/\omega \, s $, and temperatures of the hot source of $T_{h} \in [ 10^{-3} K-1 K]$, as shown in the inset.}
\label{pdf}
\end{figure}

\section{Conclusions}

We have revealed the presence of non-cyclic geometric A-A invariants in all the phenomena where
vacuum condensates appear and  shown their possible use in different physical phenomena. We have indeed shown that A-A invariant  can be utilized as novel tool  in laboratory detection of Hawking and Unruh effects, as instrument to test QFT in curved space in graphene. Also, we have suggested that a very precise quantum thermometer can be built by using geometric invariants properties.

\section*{Acknowledgement} Partial financial support from MIUR is acknowledged.

\medskip


\section*{References}


\begin{thebibliography}{99}


\bibitem{Unruh:1976db}
  Unruh W G 1976
{\it    Phys.\ Rev.\  D} {\bf 14}  870

\bibitem{Hawking:1974sw}
  Hawking S W 1975
{\it    Commun.\ Math.\ Phys.\ }  {\bf 43}  199
  [ 1976  Erratum-ibid.\  {\bf 46}  206 ].

\bibitem{Parker:1968mv}
 Parker L 1968
{\it    Phys.\ Rev.\ Lett.\ }  {\bf 21}  562

\bibitem{Schwinger:1951nm}
 Schwinger J S 1951
 {\it   Phys.\ Rev.\ } {\bf 82}  664

\bibitem{Bardeen:1957mv}
 Bardeen J, Cooper L N, Schrieffer J R 1957
 {\it   Phys.\ Rev.\   } {\bf 108}  1175
%

\bibitem{Takahasi:1974zn}
 Takahasi Y, Umezawa H 1975
{\it    Collect.\ Phenom.\ } {\bf 2}  55;
 Umezawa H 1993
  {\it  Advanced field theory: Micro, macro, and thermal physics}
(New York: AIP)

\bibitem{Blasone:1995zc}
 Blasone M, Henning P A and Vitiello G 1999
 {\it      Phys.\ Lett.\ B} {\bf 451}  140;
%
 Blasone M, Capolupo A, Vitiello G 2002
 {\it   Phys.\ Rev.\ D } {\bf 66}  025033;
 Blasone M, Capolupo A, Romei O, Vitiello G 2001
{\it    Phys.\ Rev.\ D} {\bf 63}  125015;
%
 Capolupo A, Ji C R, Mishchenko Y, Vitiello G 2004
 {\it   Phys.\ Lett.\ B} {\bf 594}  135;
 Blasone M, Capolupo A, Ji C R and Vitiello G 2010
  {\it    Int.\ J.\ Mod.\ Phys.\ A} {\bf 25}  4179;
  Blasone M, Capolupo A, Terranova F and Vitiello G 2005
  {\it    Phys.\ Rev.\ D} {\bf 72}  013003

\bibitem{Capolupo:2006et}
  Capolupo A,  Capozziello S,  Vitiello G 2007
 {\it  Phys.\ Lett.\ A} {\bf 363}, 53;
%
 2009  {\it  Phys.\ Lett.\ A} {\bf 373} 601;
%
 2008 {\it   Int.\ J.\ Mod.\ Phys.\ A} {\bf 23}, 4979;
  %
  Blasone M, Capolupo A, Capozziello S, Vitiello G 2008
{\it   Nucl.\ Instrum.\ Meth.\  A} {\bf588} 272;
%
  Blasone M, Capolupo A, Vitiello G 2010
{\it   Prog.\ Part.\ Nucl.\ Phys.\ } {\bf 64}  451;
 %
   Blasone M, Capolupo A, Capozziello S,  Carloni S, Vitiello G 2004
{\it  Phys.\ Lett.\ A}  {\bf 323}, 182

\bibitem{Capolupo:2010ek}
  Capolupo A, Di Mauro M, Iorio A 2011
 {\it   Phys.\ Lett.\  A} {\bf 375}  3415;
 %
 Capolupo  A and Di Mauro M 2012
  {\it   Phys.\ Lett.\  A} {\bf 376}, 2830;
  Capolupo  A and Di Mauro M 2013
  {\it Acta\ Phys.\ Pol. B} {\bf 44}, 81


\bibitem{Celeghini:1991yv}
 Celeghini E, Rasetti M, Vitiello G 1992
{\it    Annals Phys.\ } {\bf 215}  156

\bibitem{Iorio:2010pv}
 Iorio A 2011
 {\it   Annals Phys.\ } {\bf 326}  1334

\bibitem{Berry:1984jv}
 Berry M V 1984
 {\it  Proc.\ Roy.\ Soc.\ Lond.\  A} {\bf 392}  45


\bibitem{Anandan:1990fq}
 Anandan  J and Aharonov Y 1990
{\it   Phys.\ Rev.\ Lett.}  {\bf 65}, 1697

\bibitem{Bruno:2011xa}
  Bruno A, Capolupo A, Kak S, Raimondo G, Vitiello G 2011
  {\it Mod. Phys. Lett. B}  {\bf 25}  1661

\bibitem{Blasone:2009xk}
 Blasone M, Capolupo A, Celeghini E, Vitiello G 2009
 {\it  Phys.\ Lett.\ B} {\bf 674}  73


\bibitem{Capolupo:2011rd}
 Capolupo A 2011
 {\it     Phys.\ Rev.\ D} {\bf 84} 116002

\bibitem{Tomita}
Tomita A, Chiao R Y 1986 {\it Phys.\ Rev.\ Lett.} {\bf 57}  937


\bibitem{Jones}
Jones J A, Vedral V, Ekert A, and Castagnoli G 2000 {\it Nature}
{\bf 403}  869

\bibitem{Leek}
Leek P J, et al. 2007 {\it Science} {\bf 318}  1889
\\
Neeley M, et al. 2009 {\it Science} {\bf  325}  722

\bibitem{Pechal}
Pechal M, et al. 2012  {\it Phys.
Rev. Lett.} {\bf  108}  170401

\bibitem{Capolupo2013}
Capolupo A, Vitiello G 2013 {\it Probing Hawking and Unruh effects and quantum field theory in curved space by geometric phases,} Submitted

\bibitem{Lahav}
 Lahav O, Itah A, Blumkin A, Gordon C, Rinott S, Zayats A, and Steinhauer J 2010
{\it    Phys.\ Rev.\ Lett.} {\bf 105}  240401

\bibitem{cirac}
Garay L J, Anglin J R, Cirac J I, and Zoller P 2000
{\it    Phys.\ Rev.\ Lett.} {\bf 85}  4643

\bibitem{iorio2012}
Iorio A, Lambiase G 2012
{\it    Phys.\ Lett.\ B} {\bf 716}  334

\bibitem{Hu-Yu}
Hu J. and Yu H 2012
{\it   Phys.\ Rev.\  A} {\bf 85}  032105

\bibitem{Ivette1}
Martin-Martínez E, Fuentes I, Mann R B 2011
{\it   Phys.\ Rev.\ Lett.} {\bf 107}  131301




\end{thebibliography}
\end{document}